\documentclass[10pt,sort&compress,preprint,review]{elsarticle}
\usepackage{algorithm}
\usepackage{algorithmic}
\usepackage{bm}
\usepackage{graphicx}
\usepackage{xcolor}

\usepackage{setspace}
\usepackage{lipsum}

\usepackage[urlcolor=blue,colorlinks=true,citecolor=blue,linkcolor=blue,pdfstartview={FitH},bookmarks=false]{hyperref}

\graphicspath{{fig/}{./fig/}{.}}

\journal{Computer Physics Communications}

\begin{document}

\begin{frontmatter}

\title{GPU-based acceleration of free energy calculations in solid state physics}

\author[us]{Micha\l{} Januszewski}
\ead{michalj@gmail.com}
\author[us]{Andrzej Ptok}
\ead{aptok@mmj.pl}
\author[us]{Dawid Crivelli}
\author[us]{Bart\l{}omiej Gardas}

\address[us]{Institute of Physics, University of Silesia, 40-007 Katowice, Poland}

\begin{abstract}
Obtaining a thermodynamically accurate phase diagram through numerical calculations
is a computationally expensive problem that is crucially important to understanding
the complex phenomena of solid state physics, such as superconductivity. In this work
we show how this type of analysis can be significantly accelerated through the use of
modern GPUs. We illustrate this with a concrete example of free energy calculation
in multi-band iron-based superconductors, known to exhibit a superconducting state
with oscillating order parameter (OP). Our approach can also be used for classical
BCS-type superconductors.
With a customized algorithm and compiler tuning we are able to achieve
a 19x speedup compared to the CPU (119x compared to a single CPU core),
reducing calculation time from minutes to mere seconds,
enabling the analysis of larger systems and the elimination of finite size effects.
\end{abstract}


\end{frontmatter}

{\bf PROGRAM SUMMARY}

\begin{small}
\noindent
{\em Manuscript Title: } GPU-based acceleration of free energy calculations in solid state physics \\
{\em Authors:} Micha\l{} Januszewski, Andrzej Ptok, Dawid Crivelli, Bart\l{}omiej Gardas \\
{\em Journal Reference:}                                      \\
{\em Catalogue identifier:}                                   \\
{\em Licensing provisions: LGPLv3}                                   \\
{\em Programming language:} Fortran, CUDA C                                   \\
{\em Computer:} any with a CUDA-compliant GPU                                               \\
{\em Operating system:} no limits (tested on Linux)                                       \\
{\em RAM:} Typically tens of megabytes.                                              
{\em Classification:} 7, 6.5                                        \\
{\em Nature of problem:} GPU-accelerated free energy calculations in multi-band iron-based superconductor models.
   \\
{\em Solution method:} Parallel parameter space search for a global minimum of free energy. \\
{\em Unusual features:}\\
The same core algorithm is implemented in Fortran with OpenMP and OpenACC
compiler annotations, as well as in CUDA C. The original Fortran implementation
targets the CPU architecture, while the CUDA C version is hand-optimized for modern
GPUs.
   \\
{\em Running time:} problem-dependent, up to several seconds for a single value of momentum and
a linear lattice size on the order of $10^3$.
   \\
\end{small}
\section{Introduction}
\label{sec.intro}

The last decade brought a dynamic evolution of the computing capabilities of graphics
processing units (GPUs). In that time, the performance of a single card increased
from tens of GFLOPS in NVxx to TFLOPS in the newest Kepler/Maxwell NVIDIA chips~\cite{cuda}.
This raw processing power did not go unnoticed by the engineering and science communities,
which started applying GPUs to accelerate a wide array of calculations in what became
known as GPGPU -- general-purpose computing on GPUs. This led to the development
of special GPU variants optimized for high performance computing (e.g. the NVIDIA
Tesla line), but it should be noted that even commodity graphics cards, such as those
from the NVIDIA GeForce series, still provide enormous computational power and can
be a very economical (both from the monetary and energy consumption point of view)
alternative to large CPU clusters.

The spread of GPGPU techniques was further facilitated by the development of CUDA and OpenCL
 -- parallel programming paradigms allowing efficient exploitation of the available
GPU compute power without exposing the programmer to too many low-level details of the 
underlying hardware. GPUs were used successfully to accelerate many problems,
e.g. the numerical solution of stochastic differential equations~\cite{januszewski.kostur.09,spiechowicz.15},
fluid simulations with the lattice Boltzmann method~\cite{tolke2009,januszewski.kostur.13},
molecular dynamics simulations~\cite{anderson2008},
classical~\cite{preis2009} and quantum Monte Carlo~\cite{anderson2007} simulations, 
exact diagonalization of the Hubbard model ~\cite{siro.harju.12}, {\it etc}.

Parallel computing in general, and its realization in GPUs in particular,
can also be extremely useful in many fields of solid state physics. For
a large number of problems, the ground state of the system and
its free energy are of special interest. For instance, in order to
determine the phase diagram of a model,
free energy has to be calculated for a large number of points in the
parameter space. In this paper, we address this very issue and illustrate
it on a concrete example of a superconducting system with an oscillating
order parameter (OP), specifically an iron-based multi-band superconductor (FeSC).
Our algorithm is not limited to systems of this type and can also be used for systems
in the homogeneous superconducting state (BCS).

\begin{figure}[!h]
\begin{center}
\includegraphics{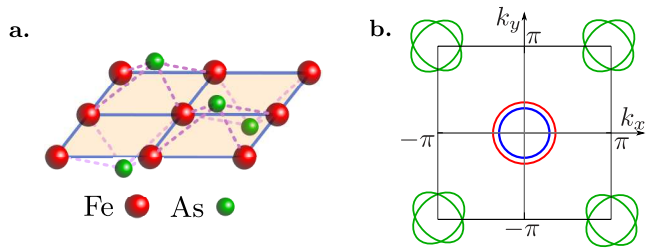}
\end{center}
\caption{(Color on-line) (Panel a) $\mathtt{FeAs}$ layers in FeSC are built by
$\mathtt{Fe}$ ions (red dots) forming a square lattice surrounded by
$\mathtt{As}$ ions (green dots) which also form a square lattice. $\mathtt{As}$
ions are placed above or under the centers of the squares formed by
$\mathtt{Fe}$. This leads to two inequivalent positions of $\mathtt{Fe}$ atoms,
so that there are two ions of $\mathtt{Fe}$ and $\mathtt{As}$ in an elementary
cell. (Panel b) True (folded) Fermi surface in the first Brillouin zone for two
$\mathtt{Fe}$ ions in unit cell. The colors blue, red and green correspond to the FS
for the 1st, 2nd, and 3rd band, respectively.} \label{fig.feas}
\end{figure}

The discovery of high temperature superconductivity in FeSC~\cite{kamihara.watanabe.08} began a
period of intense experimental and theoretical research.~\cite{ishida.nakai.09}
All FeSC include a two-dimensional structure which is shown in Fig.
\ref{fig.feas}.a. The Fermi surfaces (FS) in FeSC are composed of hole-like
Fermi pockets (around the $\Gamma = ( 0,0 )$ point) and electron-like Fermi
pockets (around the $M = ( \pi , \pi )$ point) -- Fig. \ref{fig.feas}.b.
Moreover, in FeSC we expect the presence of $s_{\pm}$ symmetry of the
superconducting OP.~\cite{mazin.singh.08} In this case the OP exhibits a sign
reversal between the hole pockets and electron pockets. For one $\mathtt{Fe}$
ion in the unit cell, the OP is proportional to $\cos k_{x}  \cdot \cos k_{y} $.

FeSC systems show complex low-energy band structures, which have been extensively studied.~\cite{mazin.singh.08,kunes.arita.10,boeri.dolgov.08,
singh.du.08,graser.maier.09,kuroki.onari.08} A consequence of this is a more
sensitive dependence of the FS to doping.~\cite{pan.li.13} In the
superconducting state, the gap is found to be on the order of 10 meV, small
relative to the breadth of the band.~\cite{ding.richard.08} This increases
the required accuracy of calculated physical quantities
needed to determine the phase diagram of the superconducting
state, such as free energy.~\cite{tai.zhu.13,ptok.14}

In this paper we show how the increased computational cost of
obtaining thermodynamically reliable results can be offset by parallelizing the
most demanding routines using CUDA, after a suitable transformation of variables
to decouple the interacting degrees of freedom. In Section \ref{sec.theory_ph} 
we discuss the theoretical background of numerical
calculations. In Section \ref{sec.algorithm} we describe the implementation of
the algorithm and compare its performance when executed on the CPU and GPU.
We summarize the results in Section \ref{sec.summary}.

\section{Theoretical background}
\label{sec.theory_ph}

Many theoretical models of FeSC systems have been proposed, with
two~\cite{raghu.qi.08},
three~\cite{daghofer.nicholson.10,daghofer.nicholson.12,korshunov.togushova.13},
four~\cite{korshunov.eremin.08} and five
bands~\cite{graser.maier.09,kuroki.onari.08}. Most of the models mentioned
describe one \texttt{Fe} unit cell and closely approximate the band and FS structure
(Fig~\ref{fig.feas}.b) obtained by LDA
calculations.~\cite{singh.du.08,ding.richard.08,kondo.santander.08,
cvetkovic.tesanovic.2.09} In every model the non-interacting tight-binding
Hamiltonian of FeSC in momentum space can be described by:
\begin{eqnarray}
H_{0} = \sum_{\alpha\beta{\bm k}\sigma} T^{\alpha\beta}_{{\bm k}\sigma} c_{\alpha{\bm k}\sigma}^{\dagger} c_{\beta{\bm k}\sigma} ,
\end{eqnarray}
where $c_{\alpha{\bm k}\sigma}^{\dagger} (c_{\alpha{\bm k}\sigma})$ is the
creation (annihilation) operator for a spin $\sigma$ electron of momentum ${\bm
k}$ in the orbital $\alpha$ (the set of orbitals is model dependent). The hopping
matrix elements $T_{{\bm k}\sigma}^{\alpha\beta} = T_{\bm k}^{\alpha\beta} -
\delta_{\alpha\beta} ( \mu + \sigma h )$ determine the model of FeSC. Here,
$\mu$ is the chemical potential and $h$ is an external magnetic field parallel
to the \texttt{FeAs} layers. For
our analysis we have chosen the minimal two-band model proposed by Raghu {\it
et al.}~\cite{raghu.qi.08} and the three-band model proposed by Daghofer {\it et
al.}~\cite{daghofer.nicholson.10,daghofer.nicholson.12} (described in 
\ref{app.twoband} and \ref{app.threeband} respectively). The band structure and FS
of the FeSC system can be reconstructed by diagonalizing the Hamiltonian $H_{0}$:
\begin{eqnarray}
H'_{0} &=& \sum_{\varepsilon{\bm k}\sigma} E_{\varepsilon{\bm k}\sigma} d_{\varepsilon{\bm k}\sigma}^{\dagger} d_{\varepsilon{\bm k}\sigma} ,
\end{eqnarray}
where $d_{\varepsilon{\bm k}\sigma}^{\dagger} (d_{\varepsilon{\bm k}\sigma})$
is the creation (annihilation) operator for a spin $\sigma$ electron of
momentum ${\bm k}$ in the band $\varepsilon$.

\paragraph{Superconductivity in multi-band iron-base systems in high magnetic fields}
FeSC superconductors are
layered~\cite{singh.du.08,ding.richard.08,kondo.santander.08,cvetkovic.tesanovic.2.09,
lioa.kondoa.09,cvetkovic.tesanovic.09}, clean~\cite{kim.tanatar.11,khim.lee.11}
materials with a relatively high Maki parameter $\alpha \sim
1-2$.~\cite{khim.lee.11,cho.kim.11,zhang.liao.11,kurita.kitagawa.11,terashima.kihou.13}
All of the features are shared with heavy fermion systems, in which strong
indications exist to observe the Fulde--Ferrell--Larkin--Ovchinnikov (FFLO)
phase~\cite{FF,LO} -- a superconducting phase with
an oscillating order parameter in real space, caused by the non-zero value of the total
momentum of Cooper pairs.

In contrast to the BCS state where Cooper pairs form a singlet state $({\bm
k}\uparrow,-{\bm k} \downarrow)$, the FFLO phase is formed by pairing states
$({\bm k}\uparrow,-{\bm k}+{\bm q} \downarrow)$. These states can occur between
the Zeeman-split parts of the Fermi surface in a high external magnetic field
(when the paramagnetic pair-breaking effects are smaller than the diamagnetic
pair-breaking effects).~\cite{matsuda.shimahara.07} In one-band materials, the
FFLO can be stabilized by anisotropies of the Fermi-surface and of the
unconventional gap function,~\cite{matsuda.izawa.06} by pair hopping
interaction~\cite{ptok.maska.09} or, in systems with nonstandard quasiparticles, with spin-dependent
mass.~\cite{kaczmarczyk.spalek.09,kaczmarczyk.spalek.10,maska.mierzejewski.10,kaczmarczyk.sadzikowski.11}
This phase can be also realized in inhomogeneous systems in the presence of
impurities~\cite{wang.hu.06,wang.hu.07,ptok.10} or spin density
waves~\cite{ptok.maska.11}. In some situations, the FFLO can be also stable in
the absence of an external magnetic field.~\cite{loder.kampf.10}
In multi-band systems, the experimental~\cite{khim.lee.11,cho.kim.11,
tarantini.gurevich.11,burger.hardy.13,zocco.grube.13} and
theoretical~\cite{ptok.14,gurevich.10,gurevich.11,ptok.crivelli.13,
mizushima.takahashi.14,takahashi.mizushima.14,crivelli.ptok.14} works point to
the existence of the FFLO phase in FeSC.
Through the analysis of the Cooper pair susceptibility in the minimal two-band
model of FeSC, such systems are shown to support the existence of an FFLO phase,
regardless of the exhibited OP symmetry. It should be noted that the state with
nonzero Cooper pair momentum, in FeSC superconductors with the $s_{\pm}$ symmetry,
is the ground state of the system near the Pauli
limit.~\cite{ptok.14,ptok.crivelli.13} This holds true also for the three-band
model (e.g. \ref{app.suscept} and Ref. \citep{crivelli.ptok.14}).

\paragraph{Free energy for intra-band superconducting phase}
In absence of inter-band interactions, the BCS and the FFLO phase (with
Cooper pairs with total momentum ${\bm q}_{\varepsilon}$ equal zero and
non-zero respectively) can be described by the effective Hamiltonian:
\begin{equation}
H_{SC} = \sum_{\varepsilon{\bm k}} \left( \Delta_{\varepsilon{\bm k}} d_{\varepsilon{\bm k}\uparrow} d_{\varepsilon,-{\bm k}+{\bm q}_{\varepsilon} \downarrow} + H.c. \right) ,
\end{equation}
where $\Delta_{\varepsilon{\bm k}} = \Delta_{\varepsilon} \eta( {\bm k} )$ is
the amplitude of the OP for Cooper pairs with total momentum ${\bm
q}_{\varepsilon}$ (in band $\varepsilon$ with symmetry described by the form factor
$\eta ({\bm k})$ -- for more details see Ref. \cite{ptok.crivelli.13}). Using
the Bogoliubov transformation we can find the eigenvalues of the full
Hamiltonian $H = H_{0} + H_{SC}$: \begin{eqnarray}
\label{eq.enequasiparticle} \mathcal{E}_{\varepsilon{\bm k}}^{\pm} &=& \frac{E_{\varepsilon{\bm k}\uparrow} - E_{\varepsilon,-{\bm k}+{\bm q}\downarrow}}{2} \pm \sqrt{ \left( \frac{E_{\varepsilon{\bm k}\uparrow} + E _{\varepsilon,-{\bm k}+{\bm q}\downarrow}}{2} \right)^{2} + | \Delta_{\varepsilon{\bm k}} |^{2} } .
\end{eqnarray}
In this case we formally describe two independent bands. The total free energy for the system is given by $\Omega = \sum_{\varepsilon} \Omega_{\varepsilon}$, where
\begin{eqnarray}
\label{eq.freeene}
\Omega_{\varepsilon} = - \frac{1}{\beta} \sum_{\alpha \in \{+, -\}} \sum_{\bm k} \ln \left( 1 +
\exp ( - \beta \mathcal{E}_{\varepsilon{\bm k}}^{\alpha} ) \right) + \sum_{\bm
k} \left( E_{\varepsilon{\bm k} \downarrow} - \frac{ 2 | \Delta_{\varepsilon}
|^{2} }{ V_{\varepsilon} } \right) , 
\end{eqnarray}
corresponding to the free energy in $\varepsilon$-th band, where $V_{\varepsilon}$ is the respective interaction intensity
and $\beta = 1 / k_B T$.

\begin{figure}[!h]
\begin{center}
\includegraphics[scale=1]{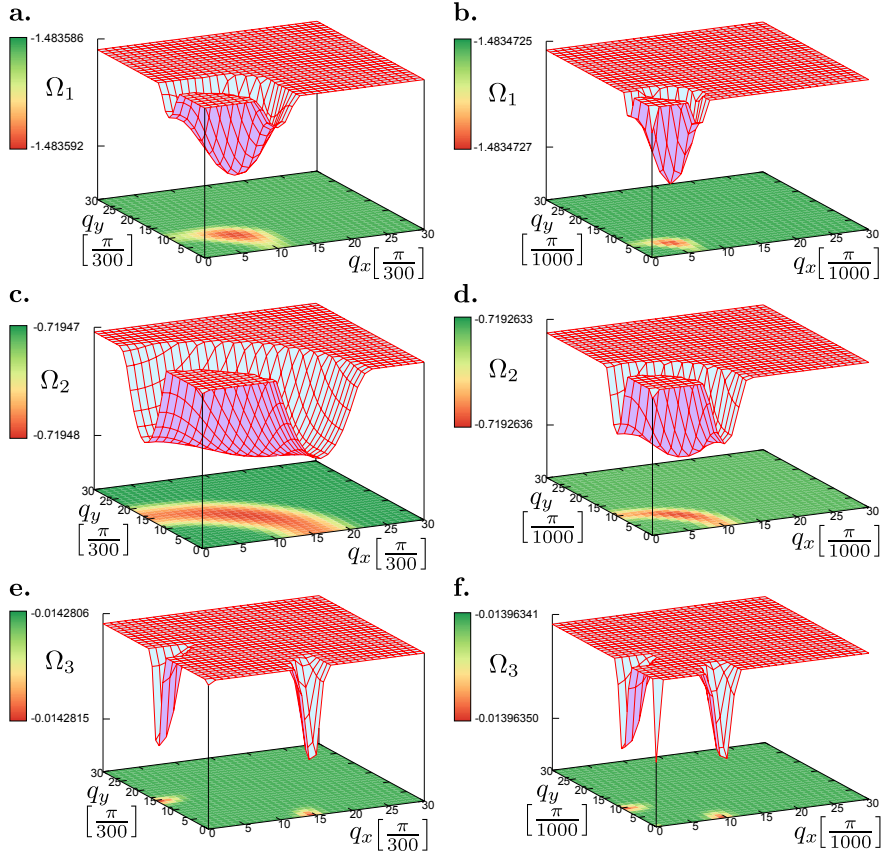}
\caption{(Color on-line) Free energy $\Omega_{\varepsilon}$ for different
parameters $V_{\varepsilon}$ in $h \simeq h_{C}^{BCS}$ -- results for
$h_{C}^{BCS} = 0.025 eV$ (panels a, c and e) and $0.005 eV$ (panels b, d and
f).} \label{fig.minene}
\end{center}
\end{figure}

\paragraph{Historical and technical note}

The historically basic concept of the FFLO phase was simultaneously proposed
by two independent groups, Fulde-Ferrell~\cite{FF} and
Larkin-Ovchinnikov~\cite{LO} in 1964. The first group proposed a
superconducting phase where Cooper pairs have only one non-zero total
momentum ${\bm q}$, and the superconducting order parameter in real space
$\Delta ( {\bm R}_{j} ) \sim \exp ( i {\bm R}_{j} \cdot {\bm q} )$. In the
second case, Cooper pairs have two possible momenta: ${\bm q}$ and the opposite
$- {\bm q}$, with an equal amplitude of the order parameter. Thus in real space
the superconducting order parameter is given by $\Delta ( {\bm R}_{j} ) \sim
\exp ( i {\bm R}_{j} \cdot {\bm q} ) + \exp ( - i {\bm R}_{j} \cdot {\bm q} ) =
\cos ( {\bm R}_{j} \cdot {\bm q} )$. However, the most general case of FFLO is
a superconducting order parameter given by a sum of plane waves, where the
Cooper pairs have all compatible values of the momentum ${\bm q}_{\alpha}$ in
the system:
\begin{equation}
\Delta ( R_{j} ) = \sum_{\alpha = 1}^{M} \Delta_{\alpha} \exp ( i {\bm R}_{j} \cdot {\bm q}_{\alpha} )
\end{equation}
where $M$ is the cardinality of the first Brillouin zone (in the square lattice
it is equal to $N_{x} \times N_{y}$). For the historical reasons described
above, whenever $M = 1$ (${\bm q}_{1} \neq 0$ and $\Delta_{1} \neq 0$) we can
speak about the Fulde--Ferrell (FF) phase, whereas for $M = 2$ (and ${\bm
q}_{1} = -{\bm q}_{2}$, $\Delta_{1} = \Delta_{2}$) about the
Larkin--Ovchinnikov (LO) phase.

Larger $M$ impose a more demanding spatial decomposition of the order
parameter, both in the theoretical and computational sense. However, every time
it can be reduced to the diagonalization of the (block) matrix representation
of the Hamiltonian. Using the translational symmetry of the lattice, the
problem for the FF phase ($M=1$) in one-band systems corresponds to the
independent diagonalization of $2 \times 2$ matrices (with eigenvalues given
like in Eq.~\ref{eq.enequasiparticle} with the number of bands $\varepsilon = 1$)
for each of the $N_{x} \times N_{y}$ different momentum sectors. In case of the
LO phase ($M=2$), the calculation can be similarly decomposed in momentum
space or using other spatial symmetries of the system (an example of this
procedure can be found in Ref.~\cite{ptok.maska.11}), with a much greater
computational effort due to the lower degree of symmetry, leading to $N_{x}$
independent diagonalization problems of size $2 N_{y} \times 2 N_{y}$. In the
{\it full} FFLO phase (i.e. in a system with
impurities~\cite{wang.hu.06,wang.hu.07,ptok.10} or a vortex lattice~\cite{maska.03}), the
spatial decomposition is determined in real space using the self-consistent
Bogoliubov-de Gennes equations, which require the full diagonalization of a
Hamiltonian of maximal rank $2 (N_x
N_y)$~\cite{loder.kampf.08,yanase.09,ptok.12,zhou.zhang.10} at every
self-consistent step. To work around these limitations, iterative
methods~\cite{litak.miller.95,martin.annett.98}  or the Kernel Polynomial
Method~\cite{weisse.wellein.06} can be used. These methods are based on the
idea of expressing functions of the energy spectrum in an orthogonal basis,
e.g. Chebyshev polynomial
expansion.~\cite{furukawa.motome.04,covaci.peeters.10,gao.huang.12,nagai.nakai.12,nagai.ota.12,nagai.shinohara.13,he.song.13}
By doing so, it becomes possible to conduct self-consistent calculations in
the superconducting state without performing the diagonalization procedure.
The time expense of iterative methods can also be reduced by a careful GPU
implementation, which is currently a work in progress.

In the present work, we describe how the calculation of the free energy can be accelerated in the FF
phase, which due to its greater symmetry allows optimal parallelization on a
GPU architecture.

\section{Parallel calculation of free energy}
\label{sec.algorithm}

\subsection{Programming models -- OpenMP, OpenACC and CUDA C}

Parallel programing can be realized in CPUs and GPUs in many different ways. In
this section we compare the performance of the same algorithm implemented
using OpenMP~\cite{openmp}, PGI CUDA/OpenACC Fortran~\cite{cloutier}, and directly in CUDA C~\cite{cuda}.

The first two are generic extensions of Fortran/C++ that make it easy to, respectively,
use multiple CPU cores, and compile a subset of existing Fortran/C++ code for a GPU.
They take the form of annotations which can be added to existing code, and as such,
enable the use of additional computational power with very little
overhead by the programmer. Typically, much better efficiency can be achieved
by the third option -- i.e. a specifically optimized implementation targeting
the GPU architecture directly. This requires more work on the part of the
programmer, both in adjusting the algorithms and in rewriting the code, but
it makes it possible to fully utilize the available resources.

\subsection{GPU algorithm}

The global ground state for a fixed magnetic field strength $h$ and temperature
$T$ is found by minimizing the free energy over the set of
$\Delta_{\varepsilon}$ and ${\bm q}_{\varepsilon}$. In case of $n$ independent
bands this corresponds to global minimization of the free energy
$\Omega_{\varepsilon}$ in every band separately, for every ${\bm
q}_{\varepsilon}$ in the first Brillouin zone (FBZ) -- Algorithm~\ref{alg.1}.

For the calculation of the free energy $\Omega_{\varepsilon}$, we must know the
eigenvalues $E_{\varepsilon{\bm k}\sigma}$ reconstructing the band structure of
our systems. In the case of the two-band model, it can simply be found analytically (see
\ref{app.twoband}). However, for models with more bands
(such as the three-band model -- \ref{app.threeband}) the band
structure has to be determined numerically (e.g. using a linear algebra
library, such as Lapack (CPU) or Magma (GPU)~\cite{magma}). With this approach,
the calculation of $E_{\varepsilon{\bm k}\uparrow}$ and $E_{\varepsilon,-{\bm
k}+{\bm q}\downarrow}$ becomes a computationally costly procedure, and if
it were to be repeated inside the inner loop of Algorithm~\ref{alg.1},
it would significantly impact the execution time. For this reason,
we propose to precalculate the eigenvalues for every momentum vector ${\bm k} \in FBZ$
and store them in memory for models with more than two bands.
The main downside of this approach is the large increase in memory usage.

While Algorithm \ref{alg.1} is simple to realize on a CPU, its execution time
is proportional to the system size $N_{x} \times N_{y}$, and as such
scales quadratically with $N_{x}$ for a square lattice ($N_{x}$
and $N_{y}$ are the number of lattice sites in the $x$ and $y$ direction,
respectively).
\begin{algorithm*}
\caption{Finding ${\bm q}_{\varepsilon}$ and $\Delta_{\varepsilon}$ corresponding to a global minimum of free energy in band $\varepsilon$.}
\label{alg.1}
\begin{algorithmic}[1]
\FOR{${\bm q}_{\varepsilon} \in FBZ$}
\STATE generate matrices  $E_{\varepsilon{\bm k}\uparrow}$ and $E_{\varepsilon,-{\bm k}+{\bm q}_{i}\downarrow}$ for ${\bm k} \in FBZ$
\FOR{$\Delta_{\varepsilon} = 0$ to $\Delta_{max}$}
\STATE calculate matrices $\mathcal{E}_{\varepsilon{\bm k}}^{\pm}$ for ${\bm k} \in FBZ$ -- Eq.~\ref{eq.enequasiparticle}
\STATE calculate $\Omega_{\varepsilon}$
\STATE find and save $\Delta_{\varepsilon}$ corresponding to a fixed ${\bm q}_{\varepsilon}$ and minimal value $\Omega_{\varepsilon}$
\ENDFOR
\STATE find and save ${\bm q}_{\varepsilon}$ and $\Delta_{\varepsilon}$ corresponding to minimum of $\Omega_{\varepsilon}$
\ENDFOR
\end{algorithmic}
\end{algorithm*}

Sometimes the physical properties of the system make it possible
to reduce the amount of computation -- for instance when it is known that
the minimum of the energy is attained for values of momentum ${\bm q}_{\varepsilon}$ in specific
directions -- Fig. \ref{fig.minene}.~\cite{ptok.14,ptok.maska.09,ptok.10,ptok.maska.11,ptok.crivelli.13,crivelli.ptok.14,
mierzejewski.ptok.10}
In this case, the outer loop of Algorithm \ref{alg.1} can be restricted to
${\bm q}_{\varepsilon} \in \mathcal{Q} \subset FBZ$, where $\mathcal{Q}$ is a set of
$N \ll N_{x} \times N_{y}$ vectors. Such reductions are not unique to linear systems
with translational symmetry but are also the case for systems with  rotational symmetry.~\cite{loder.kampf.08,yanase.09,ptok.12}

In the case of BCS-type superconductivity where Cooper pairs have zero total momentum
(${\bm q}_{\varepsilon} = 0$), Algorithm \ref{alg.1} can be further simplified
by taking into account the following property of the dispersion relation:
$E_{\varepsilon, -{\bm k}} = E_{\varepsilon {\bm k}}$ in Eq.~\ref{eq.enequasiparticle}.
This can be particularly useful in determining the system energy in the presence
of the BCS phase -- i.e. either in complete absence of external magnetic fields
 or when only weak fields are present.

A more general approach to the reduction of the execution time of our algorithm is to exploit the large degree
of parallelism inherent in the problem. In fact, Algorithm~\ref{alg.1} can be classified as
,,embarrassingly parallel'' since the vast majority of computation can be carried out independently
for all combinations of $\{{\bm q}_{\varepsilon}, {\bm k}, \Delta_{\varepsilon}\}$. For simplicity, in this paper
we concentrate on optimizing the inner loop, as all the presented methods apply to the outer
loop in a similar fashion.

We present two approaches to this problem. The first is to parallelize the execution
of the serial loop over $\Delta_{\varepsilon}$ with OpenMP to fully utilize all available CPU
cores. This has the advantage of simplicity, as the implementation requires minimal
changes to the original (serial) code.

The second approach is to implement Algorithm~\ref{alg.1} on a GPU using the CUDA
environment. Modern GPUs are capable of simultaneously executing thousands 
of threads in SIMT (Same Instruction, Multiple Threads) mode. From a programmer's
point of view, all the threads are laid out in a 1-, 2- or 3-dimensional grid and
are executing a \emph{kernel function}. The grid is further subdivided into blocks (groups
of threads), which are handled by a physical computational subunit of the GPU (the so-called
streaming multiprocessor). Threads within a block can exchange data efficiently during execution,
but cross-block communication can only take place through global GPU memory, which is significantly
slower.

\begin{figure}[!h]
\begin{center}
\includegraphics{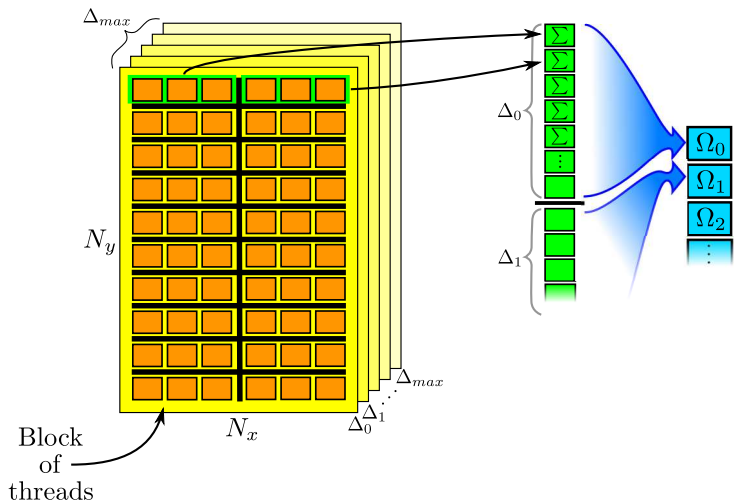}
\caption{Schematic representation Algorithm \ref{alg.1} mapped to GPU hardware.}
\label{fig.alogr}
\end{center}
\end{figure}

To fully utilize the GPU hardware, we split Algorithm~\ref{alg.1} into three steps. In the
first step, we execute the \texttt{ComputeFreeEnergy} kernel (Algorithm~\ref{alg.2}) on a
3D grid $N_x \times N_y \times \Delta_{max}$. To take advantage of the efficient intra-block
communication, we also carry out partial sums within the block (corresponding to a subset
of values spanning $N_x$) using the parallel sum-reduction algorithm.~\cite{sum}
In the second step, we execute the sum-reduction algorithm again on the partial sums
that were generated by Algorithm~\ref{alg.2}. In the third and last step, we copy
the output of step 2 from GPU memory to host memory, and look for the
value of $\Delta_i$ corresponding to the lowest free energy with a linear search. Depending on the
exact configuration of the kernels in step 1 and 2, the summation might not 
be complete at the beginning of step 3. If this is the case, we carry out
the remaining summation within the serial loop computing $\Delta_i$. With
block sizes of 128 and 1024 used for the kernels in steps 1 and 2, we can
sum up to $2^{17}$ terms in parallel on the GPU. We found that the remaining
summation was not worth the overhead of carrying it out on the GPU. Should
this not be the case for some larger problems, further parallel execution
can be trivially achieved by repeating step 2 one more time.
\begin{algorithm*}
\caption{The \texttt{ComputeFreeEnergy} CUDA kernel.}
\label{alg.2}
\begin{algorithmic}[1]
\STATE compute $\Delta_{i}$ and ${\bm k}$ corresponding to the current thread
\STATE load $E_{i{\bm k}\uparrow}$ and $E_{i,-{\bm k}+{\bm q}_{\varepsilon}\downarrow}$ from global memory (precomputed by a separate kernel)
\STATE compute $\mathcal{E}_{i{\bm k}}^{\pm}$ and $\Omega_{i}$
\STATE sum $\Omega_{i}$ for a range of ${\bm k}$ corresponding to one block of threads
\STATE save the partial sum from the previous step in global GPU memory
\end{algorithmic}
\end{algorithm*}

\subsection{Performance evaluation}

\begin{figure}[!h]
\begin{center}
\includegraphics[width=\textwidth]{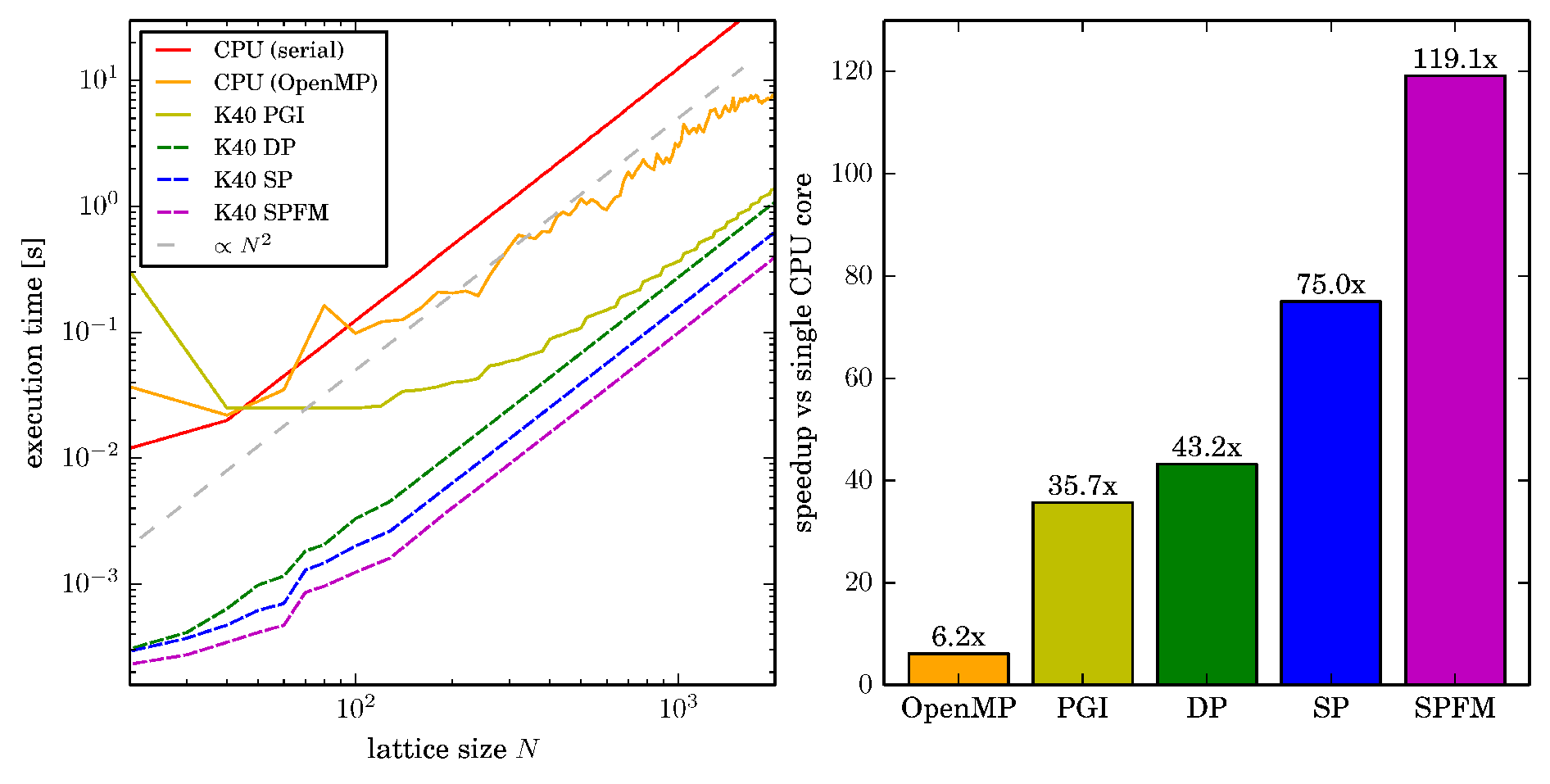}
\caption{(Color on-line) Left panel: Execution time of Algorithm~\ref{alg.1} and~\ref{alg.2} for one vector ${\bm q}_{i}$.
Right panel: speedup factors for all configurations at $N = 2000$. The last 3 case
names correspond to runs of the same CUDA C code in double precision (DP), single precision (SP),
and single precision with fast intrinsic functions (SPFM).
All versions of the Fortran code used double precision calculations.}
\label{fig.scaling}
\end{center}
\end{figure}

To test our approach, we executed Algorithms~\ref{alg.1}~and~\ref{alg.2}
on Linux machines with the following hardware:
\begin{itemize}
\item CPU: Intel(R) Core(TM) i7-3960X CPU @ 3.30GHz -- 6 cores / 12 threads,
\item GPU: NVIDIA Tesla K40 (GK180) with the SM clock set to 875 MHz.
\end{itemize}
The programs were run for a single value of ${\bm q}_i$ and 200 values
of $\Delta_i$. Calculations were done for a square lattice of size $N \times N$
for various values of $N$. The execution times (including only the computation
part of the code, and excluding any time spent on startup or input/output)
are presented in Figure~\ref{fig.scaling}.

Comparing the best CPU execution time (with OpenMP) to the GPU Fortran code using OpenACC,
we find a speedup factor of $5.8$ in the limit of large lattices.
The custom GPU code shows slightly better performance, with a $7$x speedup for the double precision
version, and additional speedup factors of $1.7$ for single precision, and $1.6$ for intrinsic
functions. When taken together,
the fastest GPU version is $19.2$ times faster than the OpenMP code and $119.1$ times faster
than the serial CPU code utilizing only a single core.

It is remarkable that the original Fortran code enhanced with OpenACC annotations provides
performance comparable to a manual implementation in CUDA C. This result shows the power
of appropriately used annotations marking parallelizable regions of the code. While still
requiring explicit input from the programmer and a good understanding of the structure of the code,
this approach is in practice significantly faster than writing the program from scratch in CUDA C
and dealing with low level details of GPU programming and resource allocation. This conclusion however only
applies in the limit of large lattices (see the left panel in Figure~\ref{fig.scaling}).
For smaller ones, the CUDA C code can be seen to be noticeably faster than OpenACC, which
is likely caused by the automatically generated GPU code introducing unnecessary overhead.

It should be noted that the last two speedup factors were achieved by trading off precision of
calculations for performance -- e.g. intrinsic functions are faster, but less precise implementations
of transcendental functions.
In our tests, we obtained the same results with all three approaches. This might not be true for
some other systems though, so we advise careful experimentation. With a factor of 2.8x between
the most and least precise method, it might also be worthwhile to run larger parameter scans
at lower precision and then selectively verify with double precision calculations.

\section{Summary}
\label{sec.summary}

The rich phenomenology and the subtle competing and interplaying phenomena of
high-$T_{C}$ materials such as FeSC (Section \ref{sec.intro}), require us to
probe fine regimes and precisely determine possible experimental signatures of
exotic phases such as FFLO (Section \ref{sec.theory_ph}).

By conducting our calculations in momentum space, and by
fully exploiting the symmetries of
the system, we are able to increase the size of the studied system by two 
orders of magnitude compared to previously reported results
and practically eliminate finite size effects.
The cost is borne by the increased complexity of the efficient
custom-tailored GPU implementation, described in Section \ref{sec.algorithm}.
Our method shown here on the example of an iron-based multi-band superconductor
exhibiting a FFLO phase, can also be used in calculations of the ground state
in standard BCS-type superconductors.

Overall, we achieved a 19x speedup compared to the CPU implementation (119x
compared a single CPU core).  In the spectrum of GPU-accelerated results in physics,
this puts us towards the higher end, with the highest speedups
being $\approx 700$x for compute-bound problems with large inherent parallelism.~\cite{januszewski.kostur.09}

\section*{Acknowledgments}

D.C. is supported by the Forszt PhD fellowship, co-funded by the European Social Fund.
B.G. is supported by the NCN project DEC-2011/01/N/ST3/02473.
The authors would like to thank NVIDIA for providing hardware resources for development and benchmarking.

\appendix

\section{Two-band model of Raghu {\it et al.}}
\label{app.twoband}

The model of FeSC proposed by Raghu {\it et al.} in Ref. \cite{raghu.qi.08}, is a minimal two-band model of iron-base pnictides describing the $d_{xz}$ and $d_{yz}$ orbitals with hybridization:
\begin{eqnarray}
T_{\bm k}^{11} &=& - 2 \left( t_{1} \cos k_{x} + t_{2} \cos k_{y} \right) - 4 t_{3} \cos k_{x} \cos k_{y} \\
T_{\bm k}^{22} &=& - 2 \left( t_{2} \cos k_{x} + t_{1} \cos k_{y} \right) - 4 t_{3} \cos k_{x} \cos k_{y} \\
T_{\bm k}^{12} &=& T_{\bm k}^{21} = - 4 t_{4} \sin k_{x} \sin k_{y} ,
\end{eqnarray}
where $t_{1} = -1.0$, $t_{2} = 1.3$, $t_{3} = -0.85$, $t_{4} = -0.85$. $|t_{1}|$ is the energy unit. Half-filling, a configuration with two electrons per site
requires $\mu = 1.54 | t_{1} |$. The model is exactly diagonalizable, with
eigenvalues: \begin{eqnarray}
E_{\pm,{\bm k}} = \frac{T_{\bm k}^{11} + T_{\bm k}^{22}}{2} \pm \sqrt{ \left( \frac{T_{\bm k}^{11} - T_{\bm k}^{22}}{2} \right)^{2}  + (T_{\bm k}^{12} )^{2} } .
\end{eqnarray}
The spectrum $E_{\alpha{\bm k}}$ reproduces the band structure and Fermi surface of FeSC -- for $\alpha = + (-)$ we get the electron-like (hole-like) band.

\begin{figure}[!b]
\begin{center}
\includegraphics[scale=0.8]{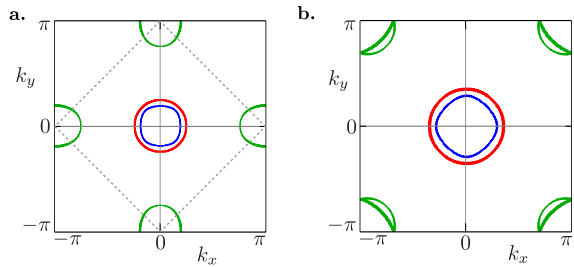}
\caption{(Color on-line) Unfolded 1Fe/cell (panel a) and folded 2Fe/cell (panel
b.) Fermi surface for the three-band model of pnictides proposed by Daghofer {\it et
al.} in Ref.~\cite{daghofer.nicholson.12}. The dashed gray line in panel a. shows the true
Brillouin zone. The colors blue, red and green correspond to the FS
for the 1st, 2nd, and 3rd band, respectively.} \label{fig.fsdag}
\end{center}
\end{figure}

\section{Three-band model Daghofer {\it et al.}}
\label{app.threeband}

This model of FeSC was proposed by Daghofer {\it et al.} in Ref. \cite{daghofer.nicholson.10} and improved in Ref. \cite{daghofer.nicholson.12}.
Beyond the $d_{xz}$ and $d_{yz}$ orbitals, the model also accounts for the $d_{xy}$ orbital:
\begin{eqnarray}
T_{\bm k}^{11} &=& 2 t_{2} \cos k_{x} + 2 t_{1} \cos k_{y} + 4 t_{3} \cos k_{x} \cos k_{y} \\
\nonumber &+& 2 t_{11} ( \cos ( 2 k_{x} ) - \cos ( 2 k_{y} ) ) + 4 t_{12} \cos ( 2 k_{x} ) \cos ( 2 k_{y} ) , \\
T_{\bm k}^{22} &=& 2 t_{1} \cos k_{x} + 2 t_{2} \cos k_{y} + 4 t_{3} \cos k_{x} \cos k_{y} \\
\nonumber &-& 2 t_{11} ( \cos ( 2 k_{x} ) - \cos ( 2 k_{y} ) ) + 4 t_{12} \cos ( 2 k_{x} ) \cos ( 2 k_{y} ) , \\
T_{\bm k}^{33} &=& \epsilon_{0} + 2 t_{5} ( \cos k_{x} + \cos k_{y} ) + 4 t_{6} \cos k_{x} \cos k_{y} \\
\nonumber &+& 2 t_{9} ( \cos ( 2 k_{x} ) + \cos ( 2 k_{y} ) ) \\
\nonumber &+& 4 t_{10} ( \cos ( 2 k_{x} ) \cos k _{y} + \cos k_{x} \cos ( 2 k_{y} ) ) , \\
T_{\bm k}^{12} &=& T_{\bm k}^{21} = 4 t_{4} \sin k_{x} \sin k_{y} , \\
T_{\bm k}^{13} &=& \bar{T}_{\bm k}^{31} = 2 i t_{7} \sin k_{x} + 4 i t_{8} \sin k_{x} \cos k_{y} , \\
T_{\bm k}^{23} &=& \bar{T}_{\bm k}^{32} = 2 i t_{7} \sin k_{y} + 4 i t_{8} \sin k_{y} \cos k_{x} .
\end{eqnarray}
In Ref. \cite{daghofer.nicholson.12} the hopping parameters in electron volts are given as: $t_{1} = -0.08$, $t_{2} = 0.1825$, $t_{3} = 0.08375$, $t_{4} = -0.03$, $t_{5} = 0.15$, $t_{6} = 0.15$, $t_{7} = -0.12$, $t_{8} = 0.06$, $t_{9} = 0.0$, $t_{10} = -0.024$, $t_{11} = -0.01$, $t_{12} = 0.0275$ and $\epsilon_{0} = 0.75$. The average number of particles in the system $n = 4$ is attained for $\mu = 0.4748$. The FS for this model is shown in Fig. \ref{fig.fsdag}.

\section{The static Cooper pair susceptibility in the three-band model}
\label{app.suscept}

The static Cooper pair susceptibility indicates the possible formation of the FFLO phase:~\cite{ptok.crivelli.13,mierzejewski.ptok.10}
\begin{eqnarray}
\chi_{\varepsilon\varepsilon'}^{\Delta} ( {\bm q} ) &\equiv & \lim_{\omega \rightarrow 0} - \frac{1}{N} \sum_{{\bm i}{\bm j}} \exp ( i {\bm q} \cdot ( {\bm i} - {\bm j} ) ) \langle \langle \widehat{\Delta}_{\varepsilon\varepsilon'{\bm i}} | \widehat{\Delta}_{\varepsilon\varepsilon'{\bm j}}^{\dagger} \rangle \rangle^{r} ,
\end{eqnarray}
where $\langle \langle . | . \rangle \rangle^{r}$ is the retarded Green's
function and $\widehat{\Delta}_{\varepsilon {\bm i}} = \sum_{\bm j} \vartheta (
{\bm j} - {\bm i} ) d_{\varepsilon {\bm i} \uparrow} d_{\varepsilon {\bm j}
\downarrow}$ is the OP in band $\varepsilon$. The operator $d_{\varepsilon{\bm
i}\sigma}$ in real space corresponds to the operator $d_{\varepsilon{\bm k}
\sigma}$ in momentum space. The factor $\vartheta ( {\bm j} - {\bm i} )$
defines the OP symmetries -- for $s_{\pm}$ pairing, $\vartheta ( {\bm \delta}
)$ is equal to $1$ for next nearest neighbors and zero
otherwise.~\cite{ptok.crivelli.13} In momentum space:
\begin{eqnarray}
\chi_{\varepsilon\varepsilon'}^{\Delta} ( {\bm q} ) = \lim_{\omega \rightarrow 0} - \frac{1}{N} \sum_{{\bm k}{\bm l}} \eta ( -{\bm k}-{\bm q} ) \eta ( -{\bm l}-{\bm q} ) \mathcal{G}_{\varepsilon\varepsilon'} ( {\bm k},{\bm l},{\bm q},\omega ) ,
\end{eqnarray}
\begin{eqnarray}
\nonumber \mathcal{G}_{\varepsilon\varepsilon'} ( {\bm k},{\bm l},{\bm q},\omega ) &=& \langle \langle d_{\varepsilon {\bm k} \uparrow} d_{\varepsilon', -{\bm k}-{\bm q} \downarrow} | d_{\varepsilon' , -{\bm l}-{\bm q} \downarrow}^{\dagger} d_{\varepsilon {\bm l} \uparrow}^{\dagger} \rangle \rangle^{r} \\
&=& \delta_{{\bm k}{\bm l}} \frac{ f ( - E_{\varepsilon {\bm k} \uparrow} ) - f ( E_{\varepsilon', -{\bm k}-{\bm q} \downarrow} ) }{\omega - E_{\varepsilon {\bm k} \uparrow} - E_{\varepsilon', -{\bm k}-{\bm q} \downarrow}} 
\end{eqnarray}
where $\eta ( {\bm k} ) = 4 \cos( k_{x} ) \cos ( k_{y} ) $ is the structure factor corresponding to the $s_{\pm}$-wave symmetry, and $f$ is the Fermi function.
This quantity can be calculated numerically similarly to the procedure used for
free energy in Section~\ref{sec.theory_ph}.

\begin{figure}[!h]
\begin{center}
\includegraphics[scale=1]{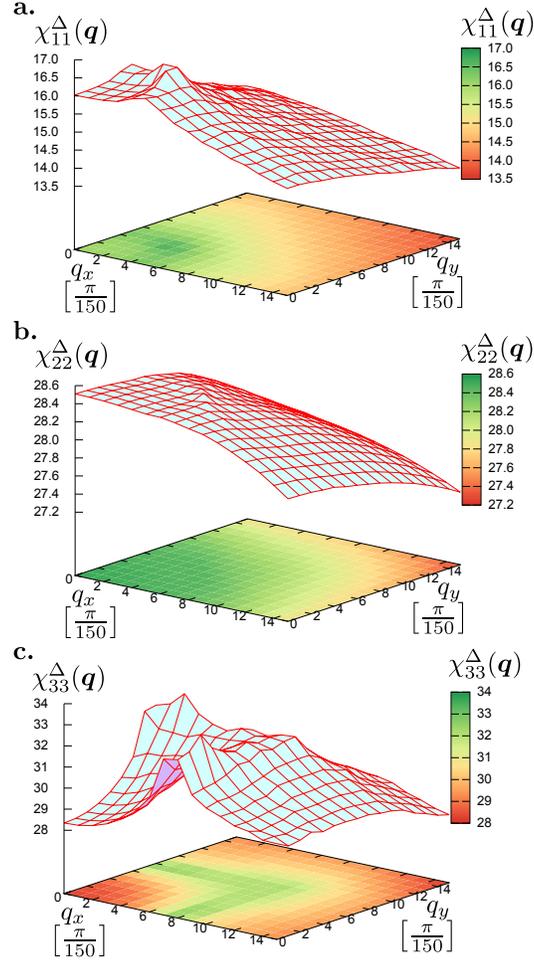}
\caption{(Color on-line) The static Cooper pair susceptibility $\chi^{\Delta}_{\varepsilon\varepsilon'}$ in magnetic field $h = 0.025$ eV and temperature $k_B T \simeq 0$ eV.}
\end{center}
\end{figure}



\begin{thebibliography}{78}

\bibitem{cuda} NVIDIA Corporation, CDUA C Programming Guide, version 5.5 (2013)

\bibitem{tolke2009} J. T\"{o}lke, M. Krafczyk, TeraFLOP computing on a desktop PC with GPUs for 3D CFD,
\href{http://dx.doi.org/10.1080/10618560802238275}{International Journal of Computational Fluid Dynamics {\bf 22} (2008) 443}

\bibitem{januszewski.kostur.13} M. Januszewski, M. Kostur, Sailfish: A flexible multi-GPU implementation of the lattice Boltzmann method,
\href{http://dx.doi.org/10.1016/j.cpc.2014.04.018}{Comput. Phys. Commun. (2014)}

\bibitem{januszewski.kostur.09} M. Januszewski, M. Kostur, Accelerating numerical solution of Stochastic Differential Equations with CUDA, 
\href{http://dx.doi.org/10.1016/j.cpc.2009.09.009}{Comput. Phys. Commun. {\bf 181} (2010) 183}

\bibitem{spiechowicz.15} J. Spiechowicz, M. Kostur, L. Machura, GPU accelerated Monte Carlo simulation of Brownian motors dynamics with CUDA,
\href{http://arxiv.org/abs/1409.4923}{Comput. Phys. Commun., accepted, 2015}

\bibitem{anderson2008} J. A. Anderson, C. D. Lorenz, A. Travesset, General purpose molecular dynamics simulations fully implemented on graphics processing units, 
\href{http://dx.doi.org/10.1016/j.jcp.2008.01.047}{J. Comput. Phys. {\bf 227} (2008) 5342}

\bibitem{preis2009} T. Preis, P. Virnau, P. Wolfgang and J. J. Schneider, GPU accelerated Monte Carlo simulation of the 2D and 3D Ising model,
\href{http://dx.doi.org/10.1016/j.jcp.2009.03.018}{J. Comput. Phys. {\bf 228} (2009) 4468}

\bibitem{anderson2007} A. G. Andersona, W. A. Goddard III, P. Schr\"{o}derb, Quantum Monte Carlo on graphical processing units, 
\href{http://dx.doi.org/10.1016/j.cpc.2007.03.004}{Comput. Phys. Commun. {\bf 177} (2007) 298}

\bibitem{siro.harju.12}T. Siro and A. Harju, Exact diagonalization of the Hubbard model on graphics processing units, 
\href{http://dx.doi.org/10.1016/j.cpc.2012.04.006}{Comput. Phys. Commun. {\bf 183} (2012) 1884}

\bibitem{kamihara.watanabe.08} Y. Kamihara, T. Watanabe, M. Hirano, H. Hosono, Iron-Based Layered Superconductor $La[O_{1-x}F_{x}]FeAs$ (x = 0.05-0.12) with $T_{c}$ = 26 K, \href{http://pubs.acs.org/doi/abs/10.1021/ja800073m}{J. Am. Chem. Soc. {\bf 130} (2008) 3296}

\bibitem{ishida.nakai.09} K. Ishida, Y. Nakai, H. Hosono, To What Extent Iron-Pnictide New Superconductors Have Been Clarified: A Progress Report, \href{http://jpsj.ipap.jp/link?JPSJ/78/062001/}{J. Phys. Soc. Jpn. {\bf 78} (2009) 062001}

\bibitem{mazin.singh.08} I.I. Mazin, D.J. Singh, M.D. Johannes, M.H. Du, Unconventional Superconductivity with a Sign Reversal in the Order Parameter of $LaFeAsO_{1−x}F_{x}$, \href{http://link.aps.org/doi/10.1103/PhysRevLett.101.057003}{Phys. Rev. Lett. {\bf 101} (2008) 057003}

\bibitem{kunes.arita.10} J. Kunes, R. Arita, P. Wissgott, A. Toschi, H. Ikeda, K. Held, Wien2wannier: From linearized augmented plane waves to maximally localized Wannier functions,
\href{http://dx.doi.org/10.1016/j.cpc.2010.08.005}{Comput. Phys. Commun. {\bf 181} (2010) 1888}

\bibitem{boeri.dolgov.08} L. Boeri, O. V. Dolgov, A. A. Golubov, Is $LaFeAsO_{1-x}F_{x}$ an Electron-Phonon Superconductor?, 
\href{http://link.aps.org/doi/10.1103/PhysRevLett.101.026403}{Phys. Rev. Lett. {\bf 101} (2008) 026403}

\bibitem{singh.du.08} D. J. Singh, M.-H. Du, Density Functional Study of $LaFeAsO_{1-x}F_{x}$: A Low Carrier Density Superconductor Near Itinerant Magnetism, \href{http://link.aps.org/doi/10.1103/PhysRevLett.100.237003}{Phys. Rev. Lett. {\bf 100} (2008) 237003}

\bibitem{graser.maier.09} S. Graser, T. A. Maier, P. J. Hirschfeld, D. J. Scalapino, Near-degeneracy of several pairing channels in multiorbital models for the Fe pnictides, \href{http://stacks.iop.org/1367-2630/11/i=2/a=025016}{New J. Phys. {\bf 11} (2009) 025016}

\bibitem{kuroki.onari.08} K. Kuroki, S. Onari, R. Arita, H. Usui, Y. Tanaka, H. Kontani, H. Aoki, Unconventional Pairing Originating from the Disconnected Fermi Surfaces of Superconducting $LaFeAsO_{1-x}F_{x}$, 
\href{http://link.aps.org/doi/10.1103/PhysRevLett.101.087004}{Phys. Rev. Lett. {\bf 101} (2008) 087004}

\bibitem{pan.li.13} L. Pan, J. Li, Y. Y. Tai, M. J. Graf, J. X. Zhu, C. S. Ting, Evolution of the Fermi surface topology in doped 122 iron pnictides, 
\href{http://link.aps.org/doi/10.1103/PhysRevB.88.214510}{Phys. Rev. B {\bf 88} (2013) 214510}

\bibitem{ding.richard.08} H. Ding, P. Richard, K. Nakayama, K. Sugawara, T. Arakane, Y. Sekiba, A. Takayama, S. Souma, T. Sato, T. Takahashi, Z. Wang, X. Dai, Z. Fang, G. F. Chen, J. L. Luo, N. L. Wang, Observation of Fermi-surface--dependent nodeless superconducting gaps in $Ba_{0.6}K_{0.4}Fe_{2}As_{2}$, 
\href{http://stacks.iop.org/0295-5075/83/i=4/a=47001}{Europhys. Lett. {\bf 83} (2008) 47001}

\bibitem{tai.zhu.13} Y. Y. Tai, J. X Zhu, M. J. Graf, C. S. Ting, Calculated phase diagram of doped BaFe2As2 superconductor in a C4-symmetry breaking model, 
\href{http://dx.doi.org/10.1209/0295-5075/103/67001}{Europhys. Lett. {\bf 103} (2013) 67001}

\bibitem{ptok.14} A. Ptok, Influence of $s_{\pm}$ symmetry on unconventional superconductivity in pnictides above the Pauli limit -- two-band model study, \href{http://dx.doi.org/10.1140/epjb/e2013-41007-2}{Eur. Phys. J. B {\bf 87} (2014) 2}

\bibitem{raghu.qi.08} S. Raghu, X. L. Qi, C. X. Liu, D. J. Scalapino, S. C. Zhang, Minimal two-band model of the superconducting iron oxypnictides, \href{http://link.aps.org/doi/10.1103/PhysRevB.77.220503}{Phys. Rev. B {\bf 77} (2008) 220503(R)}

\bibitem{daghofer.nicholson.10} M. Daghofer, A. Nicholson, A. Moreo, E. Dagotto, Three orbital model for the iron-based superconductors, \href{http://link.aps.org/doi/10.1103/PhysRevB.81.014511}{Phys. Rev. B {\bf 81} (2010) 014511}

\bibitem{daghofer.nicholson.12} M. Daghofer, A. Nicholson and A. Moreo, Spectral density in a nematic state of iron pnictides, 
\href{http://link.aps.org/doi/10.1103/PhysRevB.85.184515}{Phys. Rev. B {\bf 85} (2012) 184515}

\bibitem{korshunov.togushova.13} M. M. Korshunov, Y. N. Togushova, I. Eremin, Spin-Orbit Coupling in Fe-Based Superconductors, 
\href{http://dx.doi.org/10.1007/s10948-013-2212-6}{J. Supercond. Nov. Magn. {\bf 26} (2013) 2665}

\bibitem{korshunov.eremin.08} M. M. Korshunov, I. Eremin,Theory of magnetic excitations in iron-based layered superconductors, 
\href{http://link.aps.org/doi/10.1103/PhysRevB.78.140509}{Phys. Rev. B {\bf 78} (2008) 140509(R)}

\bibitem{kondo.santander.08} T. Kondo, A. F. Santander-Syro, O. Copie, C. Liu, M. E. Tillman, E. D. Mun, J. Schmalian, S. L. Bud'ko, M. A. Tanatar, P. C. Canfield, A. Kaminski, Momentum Dependence of the Superconducting Gap in $NdFeAsO_{0.9}F_{0.1}$ Single Crystals Measured by Angle Resolved Photoemission Spectroscopy, \href{http://link.aps.org/doi/10.1103/PhysRevLett.101.147003}{Phys. Rev. Lett. {\bf 101} (2008) 147003}

\bibitem{cvetkovic.tesanovic.2.09} V. Cvetkovic, Z. Tesanovic, Valley density-wave and multiband superconductivity in iron-based pnictide superconductors, \href{http://link.aps.org/doi/10.1103/PhysRevB.80.024512}{Phys. Rev. B {\bf 80} (2009) 024512}

\bibitem{lioa.kondoa.09} C. Liua, T. Kondoa, A.D. Palczewskia, G.D. Samolyuka, Y. Leea, M.E. Tillmana, Ni Nia, E.D. Muna, R. Gordona, A.F. Santander-Syrob, c, S.L. Bud\'{k}oa, J.L. McChesneyd, E. Rotenbergd, A.V. Fedorovd, T. Vallae, O. Copief, M.A. Tanatara, C. Martina, B.N. Harmona, P.C. Canfielda, R. Prozorova, J. Schmaliana, A. Kaminskia, Electronic properties of iron arsenic high temperature superconductors revealed by angle resolved photoemission spectroscopy (ARPES), 
\href{http://dx.doi.org/10.1016/j.physc.2009.03.050}{Physica C {\bf 469} (2009) 491}

\bibitem{cvetkovic.tesanovic.09} V. Cvetkovic, Z. Tesanovic, Multiband magnetism and superconductivity in Fe-based compounds, 
\href{http://stacks.iop.org/0295-5075/85/i=3/a=37002}{Europhys. Lett. {\bf 85}, (2009) 37002}

\bibitem{kim.tanatar.11} H. Kim, M. A. Tanatar, Y. J. Song, Y. S. Kwon, R. Prozorov, Nodeless two-gap superconducting state in single crystals of the stoichiometric iron pnictide LiFeAs, 
\href{http://link.aps.org/doi/10.1103/PhysRevB.83.100502}{Phys. Rev. B {\bf 83} (2011) 100502(R)}

\bibitem{khim.lee.11} S. Khim, B. Lee, J.W. Kim, E.S. Choi, G.R. Stewart, K.H. Kim, Pauli-limiting effects in the upper critical fields of a clean LiFeAs single crystal, 
\href{http://link.aps.org/doi/10.1103/PhysRevB.84.104502}{Phys. Rev. B {\bf 84} (2011) 104502}

\bibitem{cho.kim.11} K. Cho, H. Kim, M.A. Tanatar, Y.J. Song, Y.S. Kwon, W.A. Coniglio, C.C. Agosta, A. Gurevich, R. Prozorov, Anisotropic upper critical field and possible Fulde-Ferrel-Larkin-Ovchinnikov state in the stoichiometric pnictide superconductor LiFeAs, 
\href{http://link.aps.org/doi/10.1103/PhysRevB.83.060502}{Phys. Rev. B {\bf 83} (2011) 060502(R)}

\bibitem{zhang.liao.11} J. L. Zhang, L. Jiao, F. F. Balakirev, X. C. Wang, C. Q. Jin, and H. Q. Yuan, Upper critical field and its anisotropy in LiFeAs, 
\href{http://link.aps.org/doi/10.1103/PhysRevB.83.174506}{Phys. Rev. B {\bf 83} (2011) 174506}

\bibitem{kurita.kitagawa.11} N. Kurita, K. Kitagawa, K. Matsubayashi, A. Kismarahardja, E.-S. Choi, J. S. Brooks, Y. Uwatoko, S. Uji, T. Terashima, Determination of the Upper Critical Field of a Single Crystal LiFeAs: The Magnetic Torque Study up to 35 Tesla, 
\href{http://jpsj.ipap.jp/link?JPSJ/80/013706/}{J. Phys. Soc. Jpn. {\bf 80} (2011) 013706}

\bibitem{terashima.kihou.13} T. Terashima, K. Kihou, M. Tomita, S. Tsuchiya, N. Kikugawa, S. Ishida, C.-H. Lee, A. Iyo, H. Eisaki, S. Uji, First-order superconducting resistive transition in $Ba_{0.07}K_{0.93}Fe_{2}As_{2}$, 
\href{http://link.aps.org/doi/10.1103/PhysRevB.84.184513}{Phys. Rev. B {\bf 87} (2013) 184513}

\bibitem{FF} P. Fulde, R. A. Ferrel, Superconductivity in a Strong Spin-Exchange Field, 
\href{http://link.aps.org/doi/10.1103/PhysRev.135.A550}{Phys. Rev.  {\bf 135} (1964) A550}

\bibitem{LO} A. I. Larkin, Yu. N.Ovchinnikov, Inhomogeneous State of Superconductors, Zh. Eksp. Teor. Fiz.{\bf 47} (1964) 1136, [Sov. Phys. JETP {\bf 20} (1965) 762]

\bibitem{matsuda.shimahara.07} Y. Matsuda, H. Shimahara, Fulde-Ferrell-Larkin-Ovchinnikov State in Heavy Fermion Superconductors, 
\href{http://jpsj.ipap.jp/link?JPSJ/76/051005/}{J. Phys. Soc. Jpn. {\bf 76} (2007) 051005}

\bibitem{matsuda.izawa.06} Y. Matsuda, K. Izawa and I. Vekhter, Nodal structure of unconventional superconductors probed by angle resolved thermal transport measurements,
\href{http://dx.doi.org/10.1088/0953-8984/18/44/R01}{J. Phys.: Condens. Matter {\bf 18} (2006) R705}

\bibitem{ptok.maska.09} A. Ptok, M. M. Ma\'{s}ka, M. Mierzejewski, The Fulde-Ferrell-Larkin-Ovchinnikov phase in the presence of pair hopping interaction, \href{http://stacks.iop.org/0953-8984/21/i=29/a=295601}{J. Phys.: Condens. Matter {\bf 21} (2009) 295601}

\bibitem{kaczmarczyk.spalek.09} J. Kaczmarczyk, and J. Spa\l{}ek, Superconductivity in an almost localized Fermi liquid of quasiparticles with spin-dependent masses and effective-field induced by electron correlations,
\href{http://dx.doi.org/10.1103/PhysRevB.79.214519 }{Phys. Rev. B {\bf 79} (2009) 214519}

\bibitem{kaczmarczyk.spalek.10} J. Kaczmarczyk and J. Spa\l{}ek, Unconventional superconducting phases in a correlated two-dimensional Fermi gas of nonstandard quasiparticles: a simple model,
\href{http://dx.doi.org/10.1088/0953-8984/22/35/355702}{J. Phys.: Condens. Matter {\bf 22} (2010) 355702}

\bibitem{maska.mierzejewski.10} M. M. Ma\'{s}ka, M. Mierzejewski, J. Kaczmarczyk and J. Spa\l{}ek, Superconducting Bardeen-Cooper-Schrieffer versus Fulde-Ferrell-Larkin-Ovchinnikov states of heavy quasiparticles with spin-dependent masses and Kondo-type pairing,
\href{http://dx.doi.org/10.1103/PhysRevB.82.054509 }{Phys. Rev. B {\bf 82} (2010) 054509}

\bibitem{kaczmarczyk.sadzikowski.11} J. Kaczmarczyk, M. Sadzikowski, J. Spa\l{}ek, Conductance spectroscopy of correlated superconductor in magnetic field in the Pauli limit: evidence for strong correlations, 
\href{http://dx.doi.org/10.1103/PhysRevB.84.094525}{Phys. Rev. B {\bf 84} (2011) 094525}

\bibitem{wang.hu.06} Q. Wang, C. R. Hu, C. S. Ting, Impurity effects on the quasiparticle spectrum of the Fulde-Ferrell-Larkin-Ovchinnikov state of a d-wave superconductor
\href{http://dx.doi.org/10.1103/PhysRevB.74.212501}{Phys. Rev. B {\bf 74} (2006) 212501}

\bibitem{wang.hu.07} Q. Wang, C. R. Hu, C. S. Ting, Impurity-induced configuration-transition in the Fulde-Ferrell-Larkin-Ovchinnikov state of a d-wave superconductor
\href{http://dx.doi.org/10.1103/PhysRevB.75.184515}{Phys. Rev. B {\bf 75} (2007) 184515}

\bibitem{maska.03} M. Ma\'{s}ka, M. Mierzejewski, Vortex structure in the $d$-density-wave scenario \href{http://dx.doi.org/10.1103/PhysRevB.68.024513}{Phys. Rev. B. {\bf 68} (2003) 024513}

\bibitem{ptok.10} A. Ptok, The Fulde-Ferrell-Larkin-Ovchinnikov Superconductivity in Disordered Systems, 
\href{http://przyrbwn.icm.edu.pl/APP/ABSTR/118/a118-2-51.html}{Acta Phys. Polonica A {\bf 118} (2010) 420}

\bibitem{ptok.maska.11} A. Ptok, M. M. Ma\'{s}ka, M. Mierzejewski, Coexistence of superconductivity and incommensurate magnetic order, \href{http://link.aps.org/doi/10.1103/PhysRevB.84.094526}{Phys. Rev. B {\bf 84} (2011) 094526}

\bibitem{loder.kampf.10} F. Loder, A. P. Kampf, and T. Kopp, Superconducting state with a finite-momentum pairing mechanism in zero external magnetic field
\href{http://link.aps.org/doi/10.1103/PhysRevB.81.020511}{Phys. Rev. B 81, (2010) 020511(R)}

\bibitem{tarantini.gurevich.11} C. Tarantini, A. Gurevich, J. Jaroszynski, F. Balakirev, E. Bellingeri, I. Pallecchi, C. Ferdeghini, B. Shen, H.H. Wen, D.C. Larbalestier, Significant enhancement of upper critical fields by doping and strain in iron-based superconductors, 
\href{http://link.aps.org/doi/10.1103/PhysRevB.84.184522}{Phys. Rev. B {\bf 84} (2011) 184522}

\bibitem{burger.hardy.13} P. Burger, F. Hardy, D. Aoki, A. E. B\"{o}hmer, R. Eder, R. Heid, T. Wolf, P. Schweiss, R. Fromknecht, M. J. Jackson, C. Paulsen, C. Meingast, Strong Pauli-limiting behavior of Hc2 and uniaxial pressure dependencies in $KFe_{2}As_{2}$, 
\href{http://link.aps.org/doi/10.1103/PhysRevB.88.014517}{Phys. Rev. B {\bf 88}, (2013) 014517}

\bibitem{zocco.grube.13} D. A. Zocco, K. Grube, F. Eilers, T. Wolf, H. v. L\"{o}hneysen, Pauli-Limited Multiband Superconductivity in $KFe_{2}As_{2}$, 
\href{http://link.aps.org/doi/10.1103/PhysRevLett.111.057007}{Phys. Rev. Lett. {\bf 111}, (2013) 057007}

\bibitem{gurevich.10} A. Gurevich, Upper critical field and the Fulde-Ferrel-Larkin-Ovchinnikov transition in multiband superconductors, \href{http://link.aps.org/doi/10.1103/PhysRevB.82.184504}{Phys. Rev. B {\bf 82} (2010) 184504}

\bibitem{gurevich.11} A. Gurevich, Iron-based superconductors at high magnetic fields, 
\href{http://stacks.iop.org/0034-4885/74/i=12/a=124501}{Rep. Prog. Phys. {\bf 74} (2011) 124501}

\bibitem{ptok.crivelli.13} A. Ptok, D. Crivelli, The Fulde-Ferrell-Larkin-Ovchinnikov State in Pnictides, 
\href{http://dx.doi.org/10.1007/s10909-013-0871-0}{J. Low Temp. Phys. {\bf 172} (2013) 226}

\bibitem{mizushima.takahashi.14} T. Mizushima, M. Takahashi, K. Machida, Fulde-Ferrell-Larkin-Ovchinnikov States in Two-Band Superconductors, 
\href{http://dx.doi.org/10.7566/JPSJ.83.023703}{J. Phys. Soc. Jpn. {\bf 83} (2014) 023703}

\bibitem{takahashi.mizushima.14} M. Takahashi, T. Mizushima, K. Machida, Multiband effects on Fulde-Ferrell-Larkin-Ovchinnikov states of Pauli-limited superconductors,
\href{http://dx.doi.org/10.1103/PhysRevB.89.064505}{Phys. Rev. B {\bf 89} (2014) 064505}

\bibitem{crivelli.ptok.14} D. Crivelli, A. Ptok, Unconventional superconductivity in iron-based superconductors in a three-band model, 
\href{http://przyrbwn.icm.edu.pl/APP/ABSTR/126/a126-4a-3.html}{Acta Physica Polonica A {\bf 126} (2014) A16}


\bibitem{mierzejewski.ptok.10} M. Mierzejewski, A. Ptok, M. M. Ma\'{s}ka, Mutual enhancement of magnetism and Fulde-Ferrell-Larkin-Ovchinnikov superconductivity in $CeCoIn_{5}$, \href{http://link.aps.org/doi/10.1103/PhysRevB.80.174525}{Phys. Rev. B {\bf 80} (2010) 174525}

\bibitem{loder.kampf.08} F. Loder, A. P. Kampf, T. Kopp, Crossover from hc/e to hc/2e current oscillations in rings of s-wave superconductors, \href{http://link.aps.org/doi/10.1103/PhysRevB.78.174526}{Phys. Rev. B. {\bf 78} (2008) 174526}

\bibitem{yanase.09} Y. Yanase, Angular Fulde-Ferrell-Larkin-Ovchinnikov state in cold fermion gases in a toroidal trap, 
\href{http://link.aps.org/doi/10.1103/PhysRevB.80.220510}{Phys. Rev. B {\bf 80} (2009) 220510(R)}

\bibitem{ptok.12} A. Ptok, The Fulde-Ferrell-Larkin-Ovchinnikov State in Quantum Rings, 
\href{http://dx.doi.org/10.1007/s10948-012-1574-5}{J Supercond Nov Magn {\bf 25} (2012) 1843}

\bibitem{zhou.zhang.10} T. Zhou, D. Zhang, and C. S. Ting, Spin-density wave and asymmetry of coherence peaks in iron pnictide superconductors from a two-orbital model,
\href{http://dx.doi.org/10.1103/PhysRevB.81.052506}{Phys. Rev. B {\bf 81} (2010) 052506}


\bibitem{litak.miller.95} G. Litak, P. Miller and B.L. Gy\"{o}ffy, A recursion method for solving the Bogoliubov equations for inhomogeneous superconductors, 
\href{http://www.sciencedirect.com/science/article/pii/0921453495004386}{Phys. C {\bf 251} (1995) 263}

\bibitem{martin.annett.98} A. M. Martin and J. F. Annett, Self-consistent interface properties of d- and s-wave superconductors, 
\href{http://link.aps.org/doi/10.1103/PhysRevB.57.8709}{Phys. Rev. B {\bf 57} (1998) 8709}


\bibitem{weisse.wellein.06} A. Wei\ss{}e, G. Wellein, A. Alvermann and H. Fehske, The kernel polynomial method, 
\href{http://link.aps.org/doi/10.1103/RevModPhys.78.275}{Rev. Mod. Phys. {\bf 78} (2006) 275}


\bibitem{furukawa.motome.04} N. Furukawa and Y. Motome, Order N Monte Carlo Algorithm for Fermion Systems Coupled with Fluctuating Adiabatical Fields, 
\href{http://dx.doi.org/10.1143/JPSJ.73.1482}{J. Phys. Soc. Jpn. {\bf 73} (2004) 1482}

\bibitem{covaci.peeters.10} L. Covaci, F. M. Peeters, and M. Berciu, Efficient Numerical Approach to Inhomogeneous Superconductivity: The Chebyshev-Bogoliubov\char21{}de Gennes Method
\href{http://link.aps.org/doi/10.1103/PhysRevLett.105.167006}{Phys. Rev. Lett. {\bf 105} (2010) 167006}

\bibitem{gao.huang.12} Y. Gao, H. X. Huang and P. Q. Tong, Mixed-state effect on quasiparticle interference in iron-based superconductors,
\href{http://stacks.iop.org/0295-5075/100/i=3/a=37002}{Europhys. Lett. {\bf 100} (2012) 37002}

\bibitem{nagai.nakai.12} Y. Nagai, N. Nakai and M. Machida, Direct numerical demonstration of sign-preserving quasiparticle interference via an impurity inside a vortex core in an unconventional superconductor,
\href{http://link.aps.org/doi/10.1103/PhysRevB.85.092505}{Phys. Rev. B {\bf 85} (2012) 092505}

\bibitem{nagai.ota.12} Y. Nagai, Y. Ota and M. Machida, Efficient Numerical Self-Consistent Mean-Field Approach for Fermionic Many-Body Systems by Polynomial Expansion on Spectral Density, 
\href{http://dx.doi.org/10.1143/JPSJ.81.024710}{J. Phys. Soc. Jpn. {\bf 81} (2012) 024710}

\bibitem{nagai.shinohara.13} Y. Nagai, Y. Shinohara, Y. Futamura, Y. Ota and Sakurai, Tetsuya, Numerical Construction of a Low-Energy Effective Hamiltonian in a Self-Consistent Bogoliubov-de Gennes Approach of Superconductivity, 
\href{http://journals.jps.jp/doi/pdf/10.7566/JPSJ.82.094701}{J. Phys. Soc. Jpn. {\bf 82} (2013) 094701}

\bibitem{he.song.13} L. He and Y. Song, Self-consistent calculations of the effects of disorder in d-wave and s-wave superconductors,
\href{http://dx.doi.org/10.3938/jkps.62.2223}{J. Korean Phys. Soc. {\bf 62} (2013) 2223}

\bibitem{pgi} PGI CUDA Fortran Compiler \url{http://www.pgroup.com/resources/cudafortran.htm}

\bibitem{magma} MAGMA 1.6, (2014) \url{http://icl.eecs.utk.edu/magma}

\bibitem{sum} M. Harris, S. Sengupta, J. D. Owens, Parallel prefix sum (scan) with CUDA, GPU Gems vol. 3 (2007) 851--876


\bibitem{cloutier} B. Cloutier, B. K. Muite, and P. Rigge, Performance of FORTRAN and C GPU Extensions for a Benchmark Suite of Fourier Pseudospectral Algorithms,
\href{http://dx.doi.org/10.1109/SAAHPC.2012.24}{Proceedings of SAAHPC (2012) 145--148} 

\bibitem{openmp} OpenMP Architecture Review Board, OpenMP Application Programming Interface Version 3.0 (2008)



\end{thebibliography}
\end{document}